\documentclass[runningheads,envcountsame,envcountsect,oribibl]{llncs}

\usepackage{amsmath}
\usepackage{amssymb}
\usepackage[english]{babel}
\usepackage[short]{bibstrings}
\usepackage{general}
\usepackage{markup}
\usepackage{environment}

\title{Self-stabilizing mutual exclusion on a ring,\\
even if $K=N$\version{$Id: dijkstra.tex,v 1.5 1999/09/21 13:35:03 hoepman Exp $}}

\author{Jaap-Henk Hoepman\inst{1}}
\institute{Department of Computer Science, University of Twente, 
  the Netherlands \email{hoepman@cs.utwente.nl}}

\begin{document}

\maketitle

\bibliographystyle{alphacm-}

\begin{abstract}
We show that, contrary to common belief, Dijkstra's self-sta\-bi\-li\-zing 
mutual exclusion algorithm on a ring~\cite{Dij74,Dij82} also stabilizes when 
the number of states per node is one less than the number of nodes on 
the ring.

\noindent{\textbf{keywords}:} distributed computing, fault tolerance, 
self-stabilization.
\end{abstract}

\section{Introduction}

In~\cite{Dij74,Dij82}, Dijkstra presents the following mutual exclusion
protocol for a ring of nodes $0,\ldots,N$ where each node can read
the state $x[\cdot] \in \Set{0}{K-1}$ of its anti-clockwise neighbour, and
where node $0$ runs a different program than the other nodes.
\begin{protocol}[h]
\begin{program}
\ul* Node $0$, privileged when $x[0]=x[N]$
\ul	\IF  $x[0]=x[N]$ \THEN $x[0] \assign (x[0] + 1) \bmod K$
\ul* Node $i$, $1 \le i \le N$, privileged when $x[i] \neq x[i-1]$
\ul	\IF $x[i] \neq x[i-1]$ \THEN $x[i] \assign x[i-1]$
\end{program}
\caption{Dijkstra's mutual exclusion protocol}
\label{prot-dij}
\end{protocol}
Dijkstra proves self-stabilization of this protocol to a configuration where
only one node is privileged at a time, for $K>N$ under a central daemon and
says~\cite{Dij82}: ``for smaller values of $K$, counter examples kill the
assumption of self-stabilization''. Failing to find a counter example for 
$K=N$, we instead found the following proof that the system also stabilizes 
when $K=N$, provided that $N>1$.

\begin{theorem}
Even if $K=N$ and $N>1$,
Dijkstra's mutual exclusion protocol~\cite{Dij74,Dij82}
(Protocol~\ref{prot-dij}) stabilizes, under a central daemon, to a 
configuration where only one node is privileged.
\end{theorem}
\begin{proof}
We first define the legitimate configurations as those configurations that 
satisfy $x[i]=a$ for all $i$ with $0 \le i < j$ and $x[i]=(a-1) \bmod K$ 
for all $i$ with
$j \le i < N+1$ for some choice of $a$ and $j$. Hence the configuration where
all nodes have the same state is legitimate.

Dijkstra already showed (independent of any restriction on $K$)
closure of the legitimate states, that 
no run of the protocol ever terminates, and that in each of these runs
the exceptional node will change state (aka ``fire'') infinitely often.

Let $N>1$. Consider the case where node $0$ fires for the first time. Then just
before that, $x[0]=x[N]=b$ for some $b$ and the new value of $x[0]$ becomes
$b+1$. Now consider the case when node $0$ fires again. Then just before that,
$x[0]=x[N]=b+1$. In order for node $N$ to change value from $b$ to $b+1$, it
must have copied $b+1$ from its anti-clockwise neighbour $x[N-1]$ (which exists
if $N>1$). This moment must have occurred after node $0$ changed state to
$x[0]=b+1$. But then, just after node $N$ copies $b+1$ from node $N-1$ we
actually have $x[N-1]=x[N]=x[0]=b+1$. 

In other words, if $N>1$, three different nodes hold the same value $b+1$.
Then the remaining $N-2$ nodes can each
take a different value from the remaining $K-1$ values (unequal to $b+1$), 
which
means that if $K \ge N$ (so in particular when $K=N$) at this point in time
there is a value $a$ (among these $K-1$ values) 
not occurring as the state of any node on the ring.

Because node $0$ fires infinitely often, eventually $x[0]$ becomes $a$.
Because the other nodes merely copy values from their anti-clockwise
neighbours, at this point no other node holds $a$. The next time
node $0$ fires, $x[N]=x[0]=a$. The only way that node $N$ gets the
value $a$ is if all intermediate nodes have copied $a$ from node $0$.
We conclude that for nodes, $x[i]=a$, which is a legitimate state.
\qed
\end{proof}

\end{document}